# Single-Electron Transistor Made of a 3D Topological Insulator Nanoplate

*Yumei Jing, Shaoyun Huang\*, Jinxiong Wu, Mengmeng Meng, Xiaobo Li, Yu Zhou, Hailin Peng, Hongqi Xu\**

Dr. Y. Jing, Dr. S. Huang, Dr. M. Meng, Mr. X. Li, Prof. H. Q. Xu
Beijing Key Laboratory of Quantum Devices, Key Laboratory for the Physics and Chemistry of Nanodevices and Department of Electronics, Peking University, Beijing 100871, China
E-mails: hqxu@pku.edu.cn; syhuang@pku.edu.cn

Dr. J. Wu, Prof. H. Peng
Center for Nanochemistry, Beijing National Laboratory for Molecular Sciences (BNLMS), College of Chemistry and Molecular Engineering, Peking University, Beijing 100871, China

Prof. Y. Zhou
School of Physics and Electronics, and Institute of Super-microstructure and Ultrafast Process in Advanced Materials, Central South University, Changsha 410083, China
Powder Metallurgy Research Institute and State Key Laboratory of Powder Metallurgy, Central South University, Changsha 410083, China

Prof. H. Q. Xu
Beijing Academy of Quantum Information Sciences, Beijing 100193, China
Division of Solid State Physics, Lund University, Box 118, S-221 00 Lund, Sweden

Abstract: Quantum confined devices of three-dimensional topological insulators have been proposed to be promising and of great importance for studies of confined topological states and for applications in low energy-dissipative spintronics and quantum information processing. The absence of energy gap on the TI surface limits the experimental realization of a quantum confined system in three-dimensional topological insulators. This communication reports on the successful realization of single-electron transistor devices in $Bi_2Te_3$ nanoplates by state of the art nanofabrication techniques. Each device consists of a confined central island, two narrow constrictions that connect the central island to the source and drain, and surrounding gates. Low-temperature transport measurements demonstrate that the two narrow constrictions function as tunneling junctions and the device shows well-defined Coulomb current oscillations and Coulomb diamond shaped charge stability diagrams. This work provides a controllable and reproducible way to form quantum confined systems in three-dimensional topological insulators, which should greatly stimulate research towards confined topological states, low energy-dissipative devices and quantum information processing.

Topological insulators (TIs), which exhibit time-reversal symmetry (TRS) protected and spin-momentum inter-locked gapless Dirac surface states in the bulk band gap, are a new class of quantum matters.[1,2] Quantum confined TI devices have been proposed to be promising and of great importance for studies of the fundamental physics of confined topological states[3-8] and for applications in low energy-dissipative spintronics[9-11] and quantum information processing[12-14]. However, the absence of energy gap at the Dirac surface states and the related Klein tunneling phenomenon make it impossible to confine electrons electrostatically and to control the transport on the level of single electron via gating confinement.[15] It has been proposed and demonstrated by angle-resolved photoemission spectroscopy (ARPES) that the surface states could be gapped by magnetically doping[16] or by inter-surface state coupling[17,18]. But experimentally to exploit these effects to create quantum confined TI devices is very hard. To date, only has one experimental realization of a confined dot device in an ultrathin (7 ~ 8 nm) $Bi_2Se_3$ ribbon been reported, where the tunnel junctions were created by thinning down the ribbon locally to less than a critical thickness of 6 nm.[19] The method requires a process with an accuracy of only a few quintuple layers, considering, e.g., the fact that for $Bi_2Te_3$ the critical thickness for surface state coupling is only four quintuple-layers,[18] compromising the device process controllability. In this communication, we demonstrate the experimental realization of a 3D TI single-electron transistor device using the state of the art nanofabrication techniques. Inspired by the nanostructures etched out of graphene flakes [20-22] and of two dimensional electron gas[23,24], we have fabricated a 3D TI single-electron transistor device by carving a central island, two narrow constrictions, source and drain reservoirs, and side gates entirely out from a $Bi_2Te_3$ nanoplate via focused ion beam (FIB) technique. The fabricated device is studied by scanning probe microscope and electrical transport measurements. It is demonstrated that the fabricated device displays well-defined Coulomb blockade effect, which is the single-electron transport characteristics of a single dot, and the fabrication technique of realizing quantum confined dot in a 3D TI nanoplate has a good controllability and reproducibility. According to the fact that the Coulomb blockade effect is observed in the measurements of elctron transport through the $Bi_2Te_3$ Coulomb island, an energy gap which is essential to confine electrons on the Coulomb island in the device is expected to present on the TI surface of the two narrow constriction areas.

The TI single-electron transistor devices are fabricated from high quality $Bi_2Te_3$ nanoplates. These nanoplates are grown by van der Waals epitaxy on fluorophlogopite mica substrates and have a typical lateral size of 20 to 50 μm and a thickness of 5 to 30 nm.[25,26] After metal markers to be used in subsequent processes to locate nanoplates are defined on the growth substrates, Ti/Au (5/90 nm in thickness) contact electrodes are fabricated on top of the nanoplates by electron-beam lithography, electron-beam evaporation, and lift-off process. Narrow trenches, which define the fine structures of the devices, are then created by FIB milling. The detailed material growth and device fabrication processes are described in Supporting Information. **Figures 1A and 1B** show a schematic and a scanning electron microscope (SEM) image of a typical $Bi_2Te_3$ single-electron transistor device studied in this work, fabricated from a thin $Bi_2Te_3$ nanoplate. The device consists of a central island with an area of $A \approx 0.018$ μm² and two short and narrow constrictions with length $L \sim 25$ nm and width $W \sim 50$ nm, which connect the central island to the source and drain areas. Two central controlling side gates labeled as UG and CG, and two constriction gates labeled as LG and RG, are placed around the central island and are separated by trenches of ~50 nm in width from the central island. The contact resistances between the metal electrodes and the nanoplate have been evaluated by four-probe measurements and have been found to be negligible. For a $Bi_2Te_3$ nanoplate with a thickness of 8 ~ 30 nm, we find that the resistance between the source and drain electrodes (~ 5 μm apart) is increased by two orders of magnitude from 0.1 ~1 kΩ to 10 ~ 200 kΩ at room temperature before and after the fine structure is defined by the FIB milling. The increase in the resistance thus mainly arises from the bottleneck-like constrictions. Several fabricated devices are measured in a dilution refrigerator and they all display similar transport characteristics (Supporting Information). In the following, we concentrate to the results of our measurements for one typical device (device #14), as shown in Figure 1B. The thickness of the nanoplate, measured using atomic force microscope (AFM) is ~11 nm (Supporting Information, Figure S1), which is thick enough against coupling between the top and the bottom surface states.

**Figure 1C** shows the electron transport characteristics of the device at 40 mK. Here, the source-drain current $I_{DS}$ is measured as a function of the central gate voltage $V_{PG}$ ($V_{PG} = V_{UG} = V_{CG}$) at a source-drain voltage of $V_{DS} = 0.1$ mV and constriction gate voltages of $V_{LG} = V_{RG} = 0$ V. **Figures 1D and 1E** show the zoom-in plots of the measurements in the regions marked by two red dotted rectangles in Figure 1C. Periodic current peaks with peak-to-peak spacing $\Delta V_{PG} \approx 11$ mV are clearly observed in Figures 1D and 1E. The amplitude of the peaks is seen to be modulated in a gate-voltage scale which is an order of magnitude larger than $\Delta V_{PG}$. The peaks are reproducible with regards to different sweeping speeds of $V_{PG}$ and applied source-drain voltages $V_{DS}$ (Supporting Information, Figure S4).

**Figure 2A** shows the measured differential conductance ($dI/dV$) of the device as a function of $V_{DS}$ and $V_{PG}$ (charge stability diagram). Here, consecutive diamond-shaped regions of low conductance are observable. The horizontal width of the diamonds is equal to the current-peak separation as seen in Figures 1D and 1E. To identify the vertical size of the diamonds, we perform high resolution measurements, with the results shown in **Figure 2B**, in the area marked by a yellow dashed rectangle in Figure 2A. It is seen that all Coulomb blockade diamonds have similar vertical sizes. In addition, they are all closed at $V_{DS} = 0$ V, showing typical characteristics of single-electron transport through a single Coulomb blockaded dot in the many-electron regime. We would like to emphasize that such regular Coulomb diamonds have been observed over a wide range of gate voltages, see, e.g., Figure S5 in Supporting Information for the charge stability diagram measurements with gate voltage $V_{PG}$ being swept from −7.5 V to −6.3 V. We emphasize also that similar transport characteristics have been observed in devices made from different $Bi_2Te_3$ nanoplates with different sizes and thicknesses (Supporting Information, Figures S10 and S11), indicating that the observed Coulomb blockade effect is intrinsic to the TI island devices made by our technique. **Figure 2C** shows the measured current oscillations for the device at a source-drain voltage of $V_{DS} = 35$ μV and at temperatures of $T = 40$ to 800 mK. With increasing temperature, the peaks become broadened and the current valleys are lifted, as expected.

From the measured charge stability diagram shown in Figure 2B, the capacitance and charging energy of the $Bi_2Te_3$ island in the device can be extracted. The extracted single-electron charge energy is $E_C \approx 0.48$ meV, corresponding to a total capacitance of the dot $C_\Sigma = e^2/E_C \approx 334$ aF. The total capacitance $C_\Sigma \approx C_{PG} + C_{LG} + C_{RG} + C_S + C_D$ is a sum of all capacitances to the central island, where $C_{PG} = C_{UG} + C_{CG}$ is the capacitance of the upper side gate and the lower central side gate to the central island, $C_{LG}$ ($C_{RG}$) is the capacitance of the left (right) constriction gate to the central island, and $C_S$ ($C_D$) is the capacitance of the source (drain) to the central island. Using the capacitance formula for an isolated flat disc, $C_{disc} = 8\varepsilon_0\varepsilon_r R$,[27] where $\varepsilon_r$ is the relative permittivity of the surrounding material and $R$ is the radius of the disc, and assuming $\varepsilon_r = 80{\sim}90$ for the $Bi_2Te_3$ nanoplate,[28,29] we can extract from the measured total capacitance a value of radius $R \approx 52{\sim}59$ nm and a dot area of $A \approx 0.0085{\sim}0.011$ μm$^2$ for the central island, in good agreement with the designed area value of ~0.018 μm² of the central island.

The gate capacitances and the source and drain capacitances to the $Bi_2Te_3$ central island can also be extracted from the measured transport characteristics of the device. Based on the measured gate voltage separations of the current peaks at a small source-drain voltage, i.e., the linear response regime (Supporting Information, Figure S6), we can extract the gate capacitances as $C_{PG} \approx 15$ aF, $C_{LG} \approx 2.5$ aF and $C_{RG} \approx 3.5$ aF. The lever arm of the upper and lower central side gates is thus calculated to be $\alpha_{PG} = C_{PG}/C_\Sigma \approx 0.045$. The capacitance of the source (drain) to the central island is extracted as $C_S \approx 230$ aF ($C_D \approx 112$ aF) according to the edge slopes of the diamonds. Evidently, the total capacitance of the central island mainly arises from the source and the drain. This is reasonable because the island is electrically connected to the source and drain via tunneling.

**Figure 3A** shows the source-drain current measured for the device as a function of voltages $V_{LG}$ and $V_{RG}$ applied to the left constriction gate and the right constriction gate at source-drain voltage $V_{DS} = 0.1$ mV. Nearly horizontal (with the slope marked by a red dashed line) and vertical (with the slope marked by a black dashed line) high current stripes can be recognized. Such periodic stripes reflect single-electron populations of one single dot in the device. The two sets of independent stripes imply the presence of two small individual dots with one in each of the two constriction regions—the horizontal (vertical) current stripes correspond the presence of a small dot close to the left (right) constriction gate but far from the right (left) constriction gate. These dots are geometrically small but are tunneling-coupled to the source or the drain, and thus serve as tunnel junctions for the central island. Figure 3A also shows the traces for the existence of diagonal current stripes (with the slope marked by a yellow dashed line). To see these traces clearly, we show in **Figure 3B** a close look of the measurements in a small region marked by a red dashed rectangle in Figure 3A. Here diagonal high current stripes (again with the slope marked by a yellow dashed line) are clearly seen. Furthermore, the diagonal current lines have a slope of about −1, implying that these lines arise from the central island which couples to the left and right constriction gates nearly equally. **Figure 3C** shows the source-drain current measured as a function of voltage $V_{LG}$ applied to the left constriction gate and voltage $V_{PG}$ applied to the upper and lower central side gates at $V_{DS} = 0.1$ mV. Here, the current stripes with the slope indicated by a yellow dotted line again correspond to single-electron transport through the central island, while the current stripes with the slope indicated by a red dotted line correspond to single-electron transport through the small dot in the left constriction region. It is clearly seen that there is a capacitive coupling between the left constriction dot and the upper and lower central side gates. Similarly, **Figure 3D** shows the source-drain current measured as a function of voltage $V_{RG}$ applied to the right constriction gate and voltage $V_{PG}$ applied to the upper and lower central side gates at $V_{DS} = 0.1$ mV. Here, again, two sets of current stripes distinguished by their slopes are observed, with one set (having a slope indicated by a yellow dashed-dotted line) corresponding to single-electron tunneling through the central island and the other set (having a slope indicates by a black dashed-dotted line) corresponding to single-electron tunneling through the small dot in the right constriction region.

The measurements presented above show that a $Bi_2Te_3$ Coulomb island and two tunnel junctions that connect the island to the $Bi_2Te_3$ source and drain are formed in the $Bi_2Te_3$ nanoplate. The island is defined by the geometrical structure we have created using the FIB milling technique, as we already discussed above. To show how the two tunnel junctions could be formed in the two bottleneck-like constriction areas, we illustrate in **Figure 3E** a proposed profile of the energy bands in the device along the current direction. Here, an energy gap of the surface band is supposed to exist in the narrow constriction areas. This energy gap most probably originates from one-dimensional discrete surface subbands as a result of confinement of the surface along the perimeter direction.[30-35] Simultaneously, the subbands which are proved to be trivial show irregular fluctuations in energy due to surface roughness and disorder potentials, and the Fermi level is located in the bulk band gap (as seen from the measured resistance behavior which shows transition from a metallic property to a semiconducting property when the constriction width decreases to 50 nm, see Supporting Information, Figure S12C). When the Fermi level is located locally both in the surface and bulk band gaps, a tunneling barrier is established. In addition, as gapped surface subbands in the constriction areas are topologically trivial and could be localized by strong disorders, small dots in the constriction areas isolated out of potential fluctuations could be present.[36,37] For example, the high current stripes with the slopes marked by red and black dashed lines in Figure 3A are the results of single-electron transport through such small dots. When a small dot is at the energy degenerate point of electron numbers, single-electron tunneling could take place and the small dot becomes transparent for electrons tunneling to or out of the geometrically defined central island, leading to relatively high Coulomb current peaks. When the small dot is in the Coulomb blockade condition, the single-electron transport could still occur via co-tunneling processes through the small dot but with a less probability, leading to relatively low Coulomb current peaks. Therefore, the Coulomb oscillations observed above are intrinsic to the central island and are present in all the accessible gate voltage regions, while the amplitude of the oscillations is strongly modulated by the transparencies of the left and right constrictions. At some extremely strong blockade regions of the constrictions, the tunneling current through the device could become too low to be measurable and no Coulomb current peaks could be detected. Comparing Figure 3C to Figure 3D, we see that the left constriction dot is less well defined than the right constriction dot in the device. This difference in constriction dot definition could also be deduced from Figure 3A. Overall, as a result of disorder and surface confinement, small but transparent quantum dots could form in the bottleneck-constriction areas and the areas overall function as tunnel junctions to the central islands. It is worthwhile to mention that the lateral size of the constrictions has been found to be very critical for the constrictions to function as appropriate tunneling junctions in the realization of a $Bi_2Te_3$ single-electron transistor device (Supporting Information).

In summary, we have, in this work, demonstrated for the first time the successful realization of a 3D TI single-electron transistor device in a fully controled manner using the state of the art nanofabrication techniques. We show that a tunneling junction can be achieved by etching a nanoscale bottleneck-like constriction out of a 3D TI nanoplate. Our technique can be readily combined with top and bottom gate techniques to define single and multiple quantum dots with a tunable Fermi energy. Quantum dot systems realized in 3D TIs are desired for the study of novel physics phenomena arising from confined topological states and for the development of topological quantum computing systems.

**Experimental Section**

Experimental details and supplementary measurement results can be found in the Supporting Information.

**Supporting Information**

Supporting Information is available from the Wiley Online Library or from the authors.

**Acknowledgements**

This work was supported by the Ministry of Science and Technology of China (MOST) through the National Key Research and Development Program of China (Grant Nos. 2016YFA0300601, 2016YFA0300802, 2017YFA0303304, and 2017YFA0204901), the National Natural Science Foundation of China (Grant Nos. 11874071, 91221202, 91421303, and 11274021), and the

Beijing Academy of Quantum Information Sciences (Grant No. Y18G22). HQX also acknowledges financial support from the Swedish Research Council (VR).

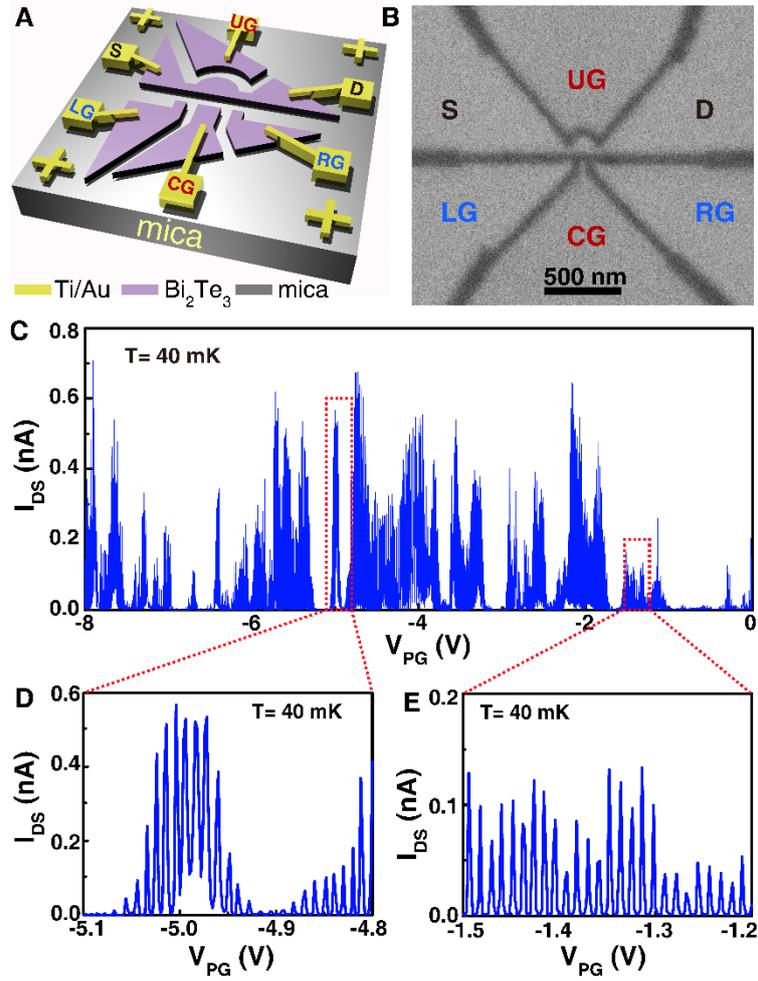

**Figure 1.** Structure and Coulomb current oscillations of a $Bi_2Te_3$ single-electron transistor device. (A) Schematic illustration of the device. The $Bi_2Te_3$ dot at the center is connected to the $Bi_2Te_3$ source and drain via two short and narrow constrictions, and is surrounded by four $Bi_2Te_3$ gates—an upper side gate (labeled by UG), a lower central side gate (labeled by CG), a left constriction gate (labeled by LG), and a right constriction gate (labeled by RG). (B) SEM image of a device measured for this work. (C) Source-drain current $I_{DS}$ measured for the device as a function of voltage $V_{PG}$ applied simultaneously to the upper side gate and the lower central side gate at source-drain voltage $V_{DS} = 0.1$ mV and voltages applied to the constriction gates $V_{LG} = V_{RG} = 0$ V. (D) and (E) Zoom-in plots of the measurements in the $V_{PG}$ ranges marked by red dotted rectangles in (C).

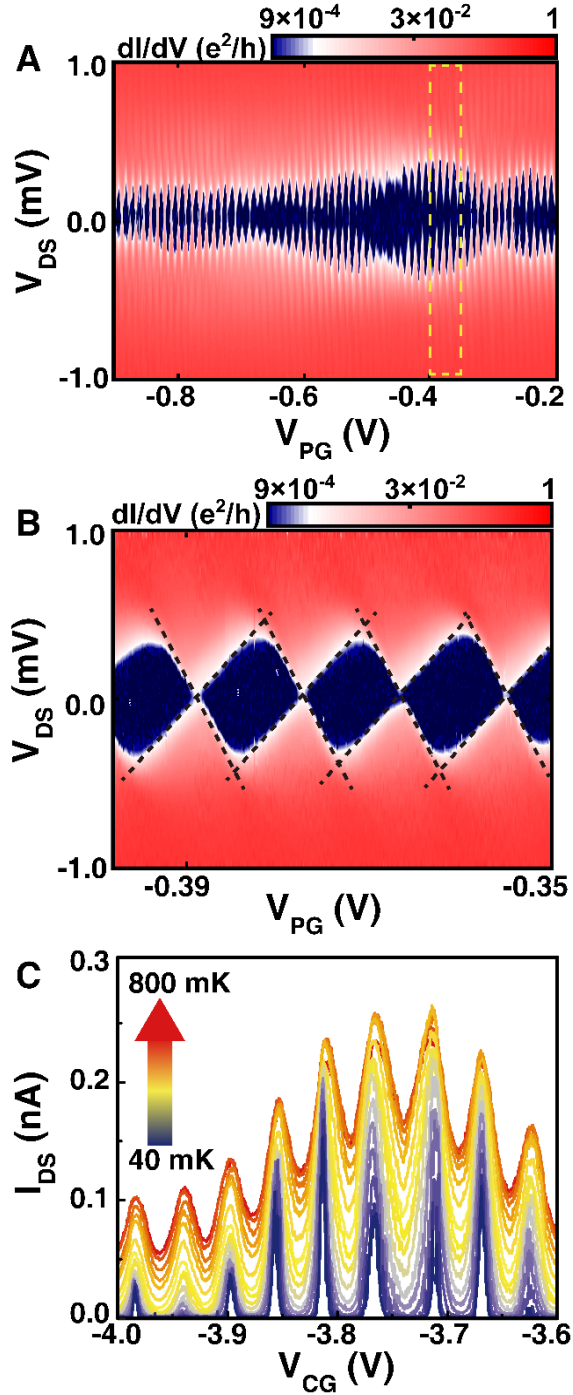

**Figure 2.** Single-electron transport characteristics of the device shown in Figure 1(B). (A) Differential conductance *dI/dV* as a function of $V_{PG}$ and $V_{DS}$ (charge stability diagram) at $T = 40$ mK and $V_{LG} = V_{RG} = 0$ V. Here, diamond shaped regions of low conductance are clearly seen. (B) High resolution measurements in the area indicated by a yellow dashed rectangle in (A). Here, regular Coulomb diamonds with roughly equal sizes can be recognized. (C) Coulomb current oscillations measured as a function of voltage $V_{CG}$ applied to the lower central gate at different temperatures at $V_{DS} = 0.035$ mV, $V_{LG} = -3.6$ V, $V_{RG} = -3$ V and $V_{UG} = -5$ V.

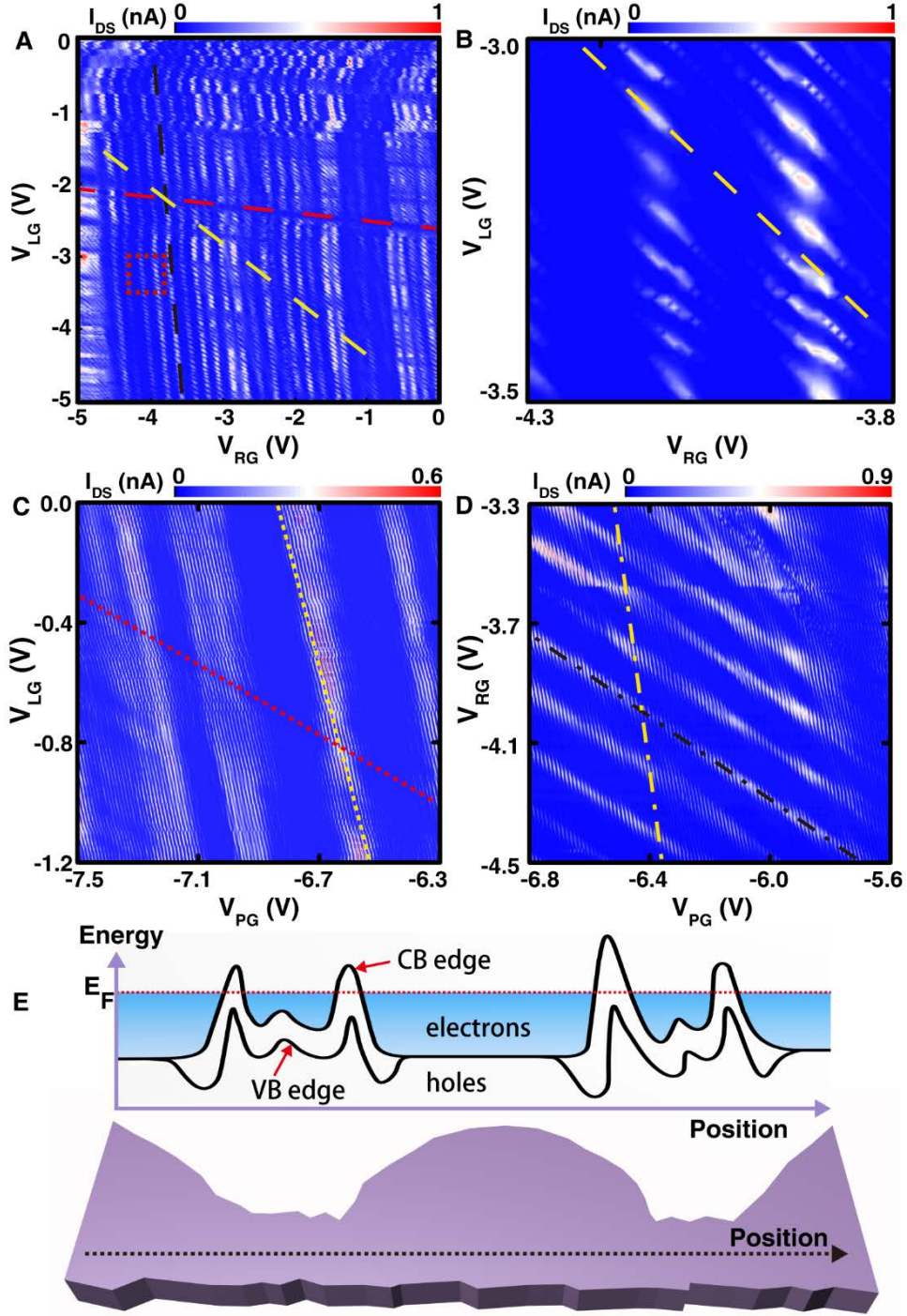

**Figure 3.** Gate voltage dependent measurements of the transport characteristics of the device shown in Figure 1(B) at source-drain voltage $V_{DS}$ = 0.1 mV. (A) Source-drain current $I_{DS}$ plotted as a function of constriction gate voltages $V_{LG}$ and $V_{RG}$ at combined upper and lower central side gate voltage $V_{PG}$ = −6.9 V. Here, nearly horizontal and nearly vertical current stripes are clearly identified. The horizontal stripes (with the slope marked by a red dashed line) indicate single-electron transport through a small dot in the left constriction. The vertical current stripes (with the slope marked by a black dashed line) reflect single-electron transport through a small dot in the right constriction. There are also signs of diagonal current stripes (with the slope marked by a yellow dashed line), arising from single-electron tunneling through the central island. (B) Highlighted results of the measurements in the area indicated by a red short-dashed rectangle in (A). Here, the diagonal current stripes are more clearly observed. (C) Source-drain current $I_{DS}$ as a function of $V_{PG}$ and $V_{LG}$ at $V_{RG} = 0$ V. The current stripes with the slope marked by a red dotted line indicate single-electron transport through the left small dot, and the current stripes with the slope marked by a yellow dotted line arise from single-electron tunneling through the central island. (D) Source-drain current $I_{DS}$ as a function of $V_{PG}$ and $V_{RG}$ at $V_{LG} = -3.6$ V. The current stripes with the slope marked by a black dash-dotted line indicate single-electron transport through the right small dot, and the current stripes with the slope marked by a yellow dash-dotted line arise from single-electron tunneling through the central island. (E) Proposed energy band profile in the device along the current direction. The upper panel shows the presence of an energy gap of surface band in the constriction areas and a dot may appear due to potential fluctuations. The upper and lower black solid curves in the constriction areas represent the lowest surface conduction band (CB) edge and the highest surface valence subband (VB) edge, while the black solid lines in the island, source and drain areas represent the Dirac points in the respective areas . To illustrate in a simple manner, the surface band gap is supposed to locate in the bulk band gap and the bulk band gap is not shown in the figure. The bottom panel shows the corresponding device structure and the position line along which the energy band profile is drawn.

# Supporting Information

### Single-Electron Transistor Made of a 3D Topological Insulator nanoplate

*Yumei Jing, Shaoyun Huang\*, Jinxiong Wu, Mengmeng Meng, Xiaobo Li, Yu Zhou, Hailin Peng,*

*and Hongqi Xu\**

**This PDF file includes:**

    Experimental Details

    Supplementary Measurement Results of the Device Studied in the Main Article

    Measurements of Additional $Bi_2Te_3$ Single-Electron Transistor Devices

    Measurements of a Single Constriction

    Figures S1 to S12

    Table S1

    References

\*Correspondence should be addressed to Prof. Hongqi Xu (hqxu@pku.edu.cn) or Dr. Shaoyun Huang (syhuang@pku.edu.cn).

## Experimental Details

Materials

Bi$_2$Te$_3$ nanoplates are grown via van der Waals epitaxy on the atomically smooth surfaces of mica substrates.[1,2] During the epitaxial growth processes, Bi$_2$Te$_3$ first nucleates at random locations on the mica surfaces and then grows into triangular or hexagonal nanoplates. These nanoplates can grow up to a lateral size of 20 ~ 50 μm and to a thickness from few to tens of quintuple layers by tuning growth parameters. High quality of the nanoplates has previously been demonstrated using microstructure analysis tools, including transmission-electron microscope (TEM).[1] Hall measurements show that as-grown Bi$_2$Te$_3$ nanoplates are n-type materials and have an electron concentration of ~$5 \times 10^{13}$ cm$^{-2}$ at 2 K.

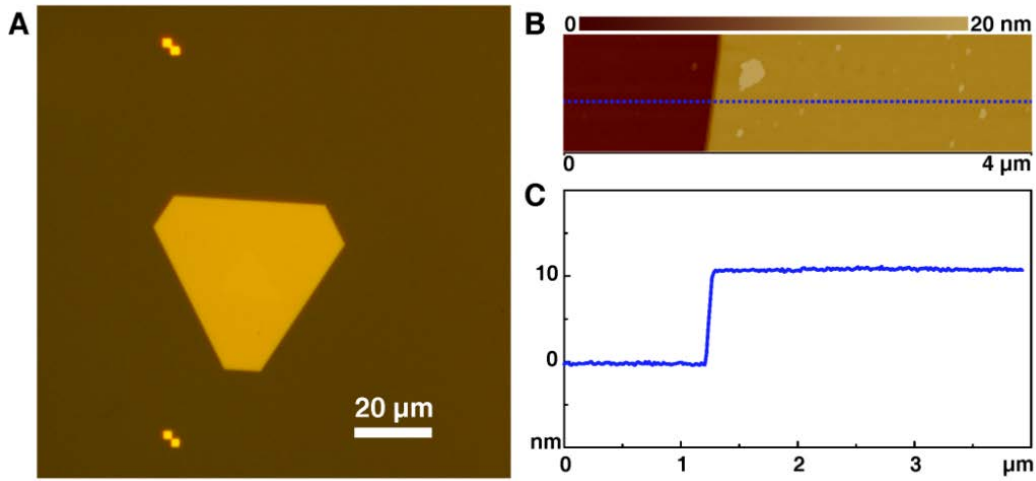

**Figure S1.** Morphology characterization of as-grown Bi$_2$Te$_3$ nanoplates. (A) Optical microscope image of a nanoplate on the growth substrate used for fabrication of device #14 reported in the main article after the metal alignment markers are fabricated on the substrate. The nanoplate is in a hexagonal shape with a size of ~40 μm measured as the distance between two parallel sides. (B) AFM image of the nanoplate. Particles seen on the nanoplate are contaminations introduced in the marker fabrication process. (C) Cross-sectional analysis of the nanoplate carried out along the dashed blue line in (B). The analysis shows that the nanoplate has a thickness of ~11 nm.

Figure S1 shows, as an example, the morphology analysis of the nanoplate used for fabricating the Bi$_2$Te$_3$ single-electron transistor device (device #14) reported in the main article, after fabrication of metal alignment markers on the growth substrate. The optical microscope image displayed in Figure S1A shows that the nanoplate is of a hexagonal shape with a size (measured as a distance between two parallel sides) of ~40 μm. The atomically smooth surface has been identified by measurements using an atomic force microscope (AFM) shown in Figure S1B. The particles seen on the nanoplate are contaminations introduced in the marker fabrication process. The thickness of the nanoplate is found to be ~11 nm from the AFM cross-sectional analysis shown in Figure S1C.

Device fabrication

The $Bi_2Te_3$ single-electron transistor devices studied in the main article and the Supporting Information are fabricated in a process shown schematically in Figure S2. In order to avoid damages of as-grown $Bi_2Te_3$ nanoplates introduced by any transfer process, all the devices are fabricated directly on the growth insulating mica substrates [Figure S2A]. After a mica substrate with as-grown $Bi_2Te_3$ nanoplates on it is selected, small alignment markers are prepared by electron-beam lithography (EBL), metal evaporation and lift-off techniques, in order to locate $Bi_2Te_3$ nanoplates for subsequent device fabrication processes [Figure S2B]. Here, a technology challenge due to charge accumulation arises for the EBL process on the insulating mica substrate. In order to solve the charging-up problem, after spin coating of electron-beam resist (PMMA) on the mica substrate, we have coated a thin layer of conducting polymer (SX AR-PC 5000/90.1) on top of PMMA. Then, the same EBL, metal evaporation, and lift-off processes are employed to fabricate Ti/Au (5/90 nm in thickness) electrodes, which contact to selected nanoplates, on the mica substrate [Figure S2C]. Here, we note that in order to achieve good ohmic contacts to $Bi_2Te_3$ nanoplates, argon plasma treatment is applied on the contact areas right before the metal evaporation process. We note also that the Ti/Au metal leads, which connect the contact electrodes to the sample holder ground, are fabricated in the same time when the contact electrodes are fabricated, in order to eliminate the charging-up effect in the subsequent focus ion-beam (FIB) process. The Coulomb island structures are then defined via FIB milling, which directly writes trenches with designed patterns on the selected nanoplates in a high resolution without a need of mask [Figure S2D]. Finally, the grounding leads are cut by mechanically scraping to disconnect the electrodes of the devices from the sample holder ground.

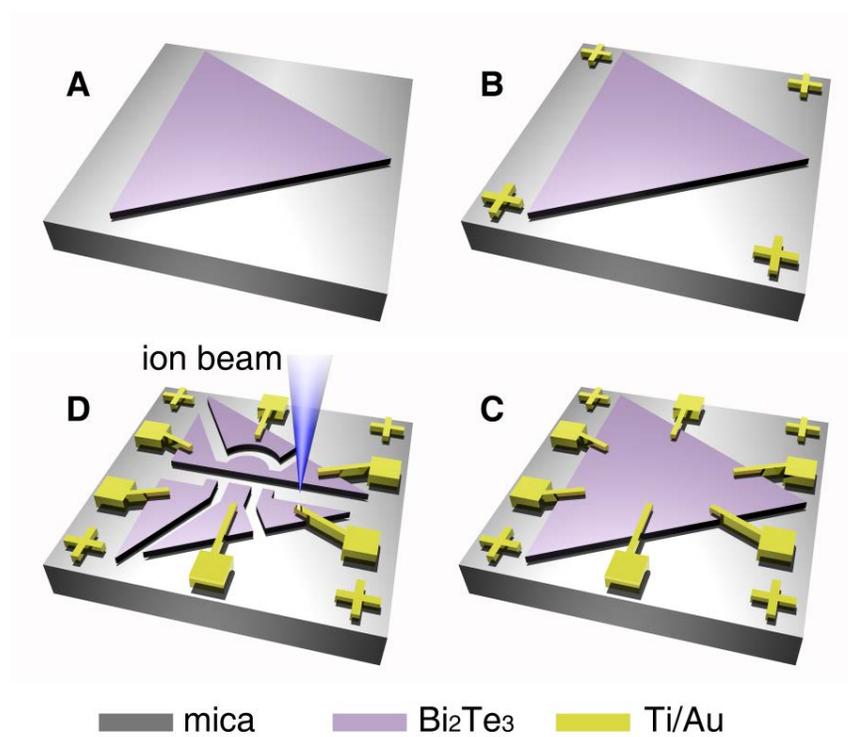

**Figure S2.** Schematic illustration of the fabrication process of $Bi_2Te_3$ single-electron transistor devices on the growth substrates of mica. (A) As-grown $Bi_2Te_3$ nanoplate on top of a growth mica substrate selected for device fabrication. (B) Preparation of metal alignment markers using standard electron-beam lithography, metal evaporation and lift-off processes. (C) Fabrication of electrodes and grounding leads (the latter are not shown) with the same processes as (B). (D) Carving a Coulomb island and surrounding gates out of the $Bi_2Te_3$ nanoplate with focus ion-beam milling.

Structure design

Devices with different constriction sizes have been designed and fabricated to find out an optimal size of constrictions. Table S1 lists the room temperature source-drain resistances of the testing devices made on $Bi_2Te_3$ nanoplates with different thicknesses and different constriction widths, but similar constriction lengths. In general, a too long constriction often leads to formation of a series of quantum dots in the constriction. In Table S1, all the devices are designed to be 20 ~ 25 nm in constriction length. A too wide constriction has too low impedance to form an effective tunneling barrier required for defining a Coulomb island, whereas a too narrow constriction exhibits too high impedance to measure the transport current. In Table S1, devices with a room temperature resistance less than ~10 kΩ do not show any Coulomb blockade behavior at low temperature, while devices with a room temperature resistance higher than ~100 kΩ are always found to show the characteristics of multiple QDs. For nanoplates with a thickness of ~10 nm, the optimal constriction size to establish an effective quantum barrier in a constriction is ~25 nm in length and ~50 nm in width and the devices made with these optimized constrictions exhibit a source-drain resistance of 10 to 100 kΩ at room temperature. As the resistance is nanoplate thickness dependent, the constriction width of the devices made on $Bi_2Te_3$ nanoplates with different thicknesses deviate from the optimal parameters for the devices made on the nanoplates with a thickness of ~10 nm.

**Table S1.** Room temperature source-drain resistance of testing single-electron transistor devices made from $Bi_2Te_3$ nanoplates with different thicknesses and constriction widths. All the devices are designed to be 20 ~ 25 nm in constriction length.

| Thickness (nm) | Constriction width (nm) | Resistance (kΩ) | Thickness (nm) | Constriction width (nm) | Resistance (kΩ) |
|---|---|---|---|---|---|
| 8 | 50 | 90 | 14 | 40 | 30 |
| 8 | 40 | 50 | 15 | 40 | 12 |
| 8 | 40 | 310 | 15 | 60 | 3 |
| 8 | 50 | 78 | 15 | 40 | 20 |
| 9 | 50 | 40 | 16 | 40 | 20 |
| 10 | 50 | 10 | 17 | 60 | 10 |
| 11 | 50 | 20 | 17 | 50 | 10 |
| 12 | 50 | 25 | 17 | 40 | 18 |
| 12 | 30 | 230 | 18 | 30 | 10 |
| 13 | 40 | 15 | 20 | 30 | 14 |
| 13 | 40 | 48 | 34 | 50 | 30 |

Low-temperature Transport Measurements

A few fabricated single-electron transistor devices with optimally designed structures are selected for low-temperature transport measurements. All the low-temperature transport measurements are performed in a $^3He/^4He$ dilution refrigerator. In order to avoid heating-up the electron temperature in the devices by electrical noise of the measurement environment, a cooper powder filter and a RC filter are employed in the measurement circuit at the milli-Kelvin cold end. In the measurements, a DC voltage is applied to the source with the drain being grounded, a DC current is record using a current preamplifier (DL-1211). The differential conductance is obtained by numerical calculations from the measured I-V curves. All the other noninvolved electrodes in a measurement are grounded if not specifically noted.

## Supplementary Measurement Results of the Device Studied in the Main Article

Characteristic details and stability of the Coulomb current oscillations

As we showed in Figure 1 of the main article, Coulomb oscillations seen in the measured current as a function of $V_{PG}$ are widely observed. These oscillations have roughly equal separation in gate voltage $V_{PG}$. Here, as an additional example, we show in Figure S3 the current measured for the same device (Device # 14) as a function of $V_{PG}$ at the source-drain bias voltage $V_{DS} = 0.2$ mV. Again, rather equally $V_{PG}$-spaced Coulomb oscillations are observed. In addition, here it is seen that the current can reach zero between peaks in a wide range of $V_{PG}$.

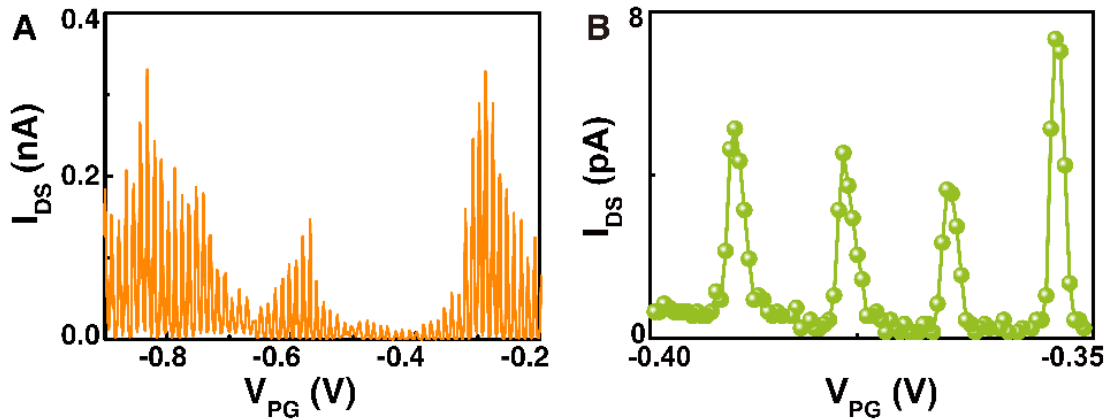

**Figure S3.** Measured current oscillations of the $Bi_2Te_3$ single-electron transistor device (device #14) studied in the main article. (A) Source-drain current $I_{DS}$ of the device measured as a function of $V_{PG}$ ($\boldsymbol{V_{PG}} = \boldsymbol{V_{UG}} = \boldsymbol{V_{CG}}$) in the voltage range of -0.9 to -0.2 V at $V_{DS} = 0.2$ mV, $V_{LG} = V_{RG} = 0$ V and $T = 40$ mK. (B) Zoom-in plot of the measurements in the $V_{PG}$ range of -0.4 to -0.35 V.

The observed characteristics of the Coulomb current oscillations shown in the main article are robust and reproducible against gate sweeping speed and applied source-drain voltage. Figure S4 shows the source-drain current measured as a function of voltage $V_{PG}$ applied to the central and upper gates with (A) different $V_{PG}$ sweeping steps and (B) different applied source-drain voltages $V_{DS}$. Here, it is seen that the current oscillations remain the same in both peak position and peak amplitude in the measurements with two different $V_{PG}$ sweeping steps. It is also seen that when $V_{DS}$ decreases, the current peak positions remain unchanged, but the peak widths become narrowed as expected.

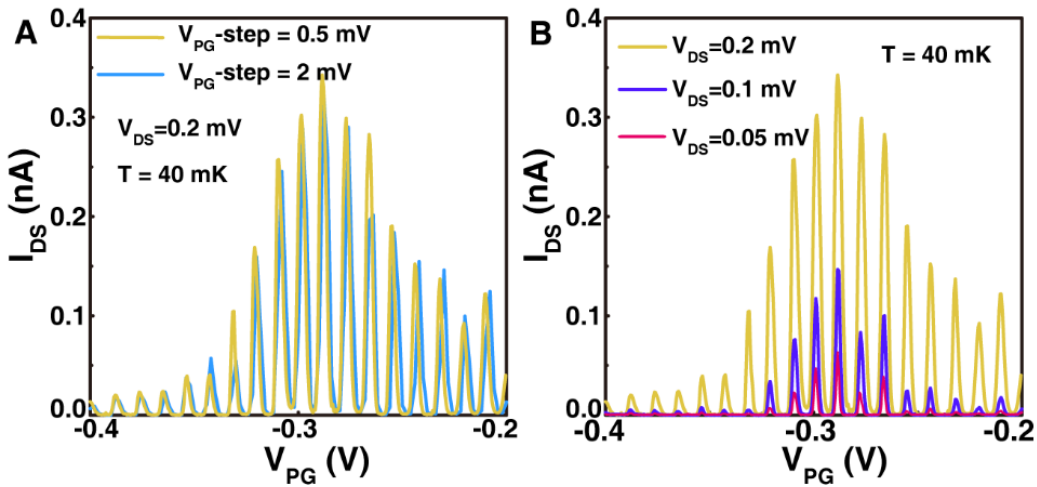

**Figure S4.** Reproducibility of the measurement results. (A) Source-drain current $I_{DS}$ of the device (device #14) measured as a function of voltage $V_{PG}$ applied to the central and upper gates at $V_{DS} = 0.2$ mV, $V_{LG} = V_{RG} = 0$ V and $T = 40$ mK with different $V_{PG}$ sweeping steps. (B) Source-drain current $I_{DS}$ measured as a function of $V_{PG}$ at $V_{LG} = V_{RG} = 0$ V, $T = 40$ mK and at three different source-drain voltages $V_{DS}$.

Additional measured charge stability diagram

The characteristics seen in the charge stability diagram of the $Bi_2Te_3$ single-electron transistor device (Device #14) shown in Figure 2 of the main article are also seen in a wide range of gate voltage $V_{PG}$. Figure S5A,B shows the measured Coulomb current oscillations and the charge stability diagram of the device in a more negative range of gate voltage $V_{PG}$. Here, it is seen that the charge stability diagram of the device shows the same characteristics as seen in Figure 2A of the main article.

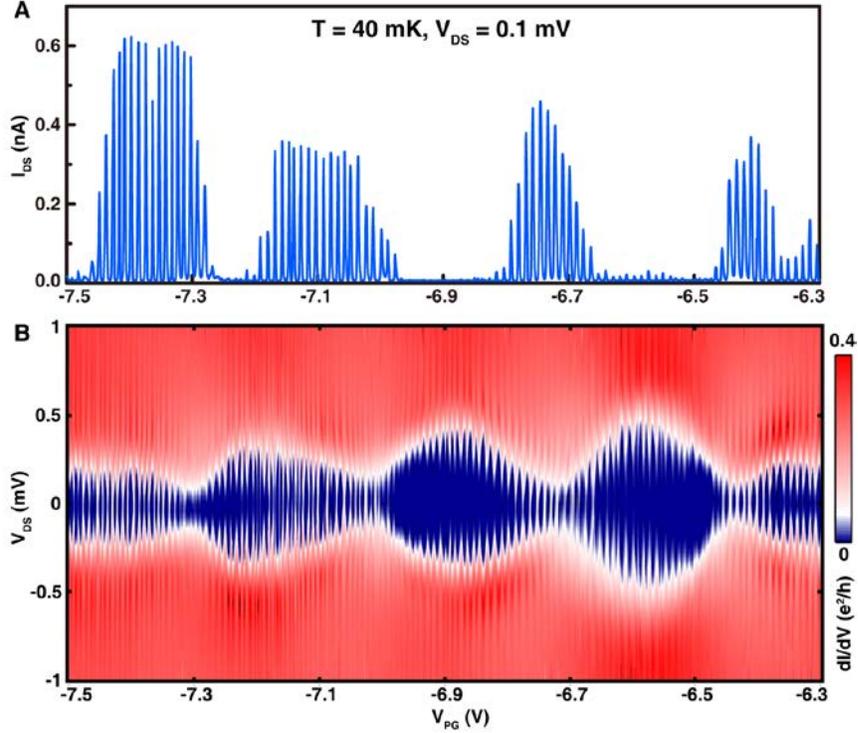

**Figure S5.** Electron transport properties of the $Bi_2Te_3$ single-electron transistor device (device #14) studied in the main article in a more negative $V_{PG}$ region. (A) Source-drain current $I_{DS}$ as a function of $V_{PG}$ in the range of -7.5 to -6.3 V measured at $V_{DS}$ = 0.1 mV, $V_{LG}$ = $V_{RG}$ = 0 V and $T$ = 40 mK. (B) Differential conductance of the device measured as a function of $V_{PG}$ and $V_{DS}$ (charge stability diagram) in the same $V_{PG}$ range at $T$ = 40 mK.

Functionality and tunability of different side gates

In the $Bi_2Te_3$ single-electron transistor device, every lateral gate is found to be capable of coupling to the central island effectively. Figure S6A-C shows the current measured as a function of voltage $V_{LG}$ applied the left constriction gate, voltage $V_{RG}$ applied to the right constriction gate, and voltage $V_{PG}$ applied to the two middle gates (i.e., the central and upper gates), respectively. It is seen that all the side gates can individually modulate the electron number in the island and thus the current through the device effectively. Nevertheless, different lateral gates do exhibit different values of the coupling capacitance to the central island. Figure S6D summarizes the current peak spacing in gate voltage ($\Delta V_G$) extracted from the Coulomb current oscillations shown in Figure. S6A-C. The gate capacitance to the central island can be calculated as $e/\Delta V_G$. Clearly, the Coulomb current oscillations with regard to $V_{PG}$ have the smallest period, while the periods of the oscillations in $V_{LG}$ and $V_{RG}$ are relatively much larger and roughly equal to each other. Therefore, the central and upper gates exhibit a stronger capacitive coupling to the central island than the left and right constriction gates.

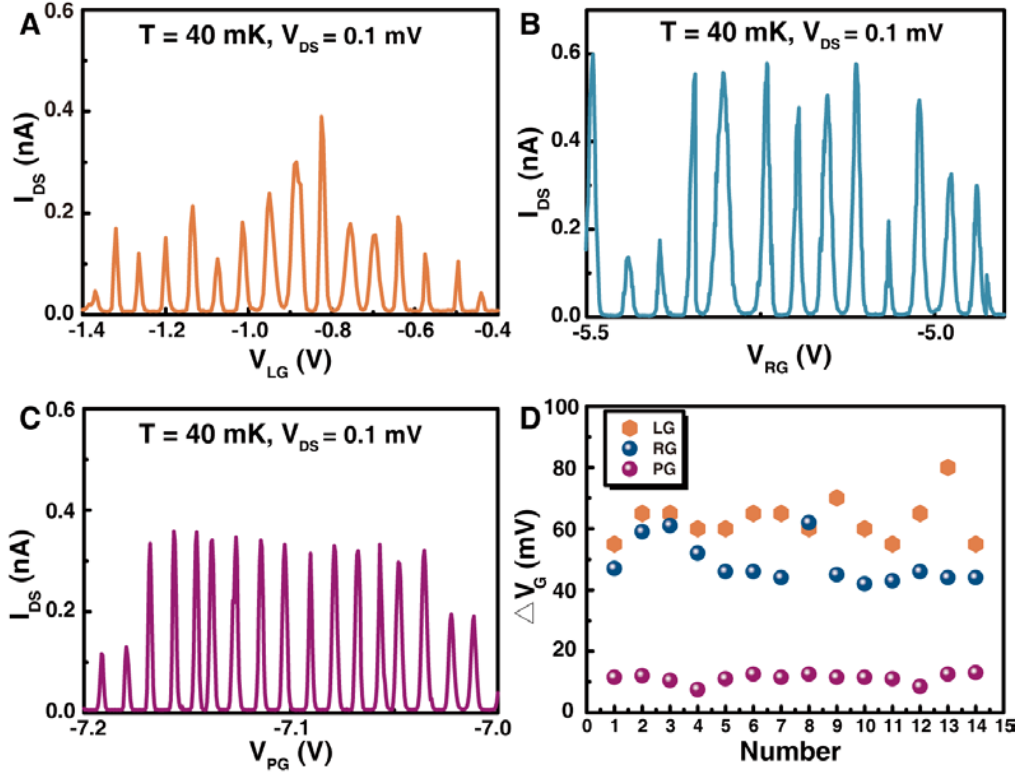

**Figure S6.** Gate voltage dependent measurements of the Coulomb current oscillations of the device (device #14) studied in the main article. (A) Source-drain current of the device measured as a function of voltage $V_{LG}$ applied to the left constriction gate at voltage applied to the central and upper gates $V_{PG}$ = -6.3 V, voltage applied to the right constriction gate $V_{RG}$=0 V, source-drain bias voltage $V_{DS}$=0.1 mV, and $T$ = 40 mK. (B) Source-drain current of the device measured as a function of voltage $V_{RG}$ applied to the right constriction gate at $V_{PG}$ = -6.7 V, $V_{LG}$= -3.6 V, $V_{DS}$=0.1 mV and $T$ = 40 mK. (C) Source-drain current of the device measured as a function of voltage $V_{PG}$ applied to the central and upper gates at $V_{LG}$= $V_{RG}$= 0 V, $V_{DS}$=0.1 mV and $T$ = 40 mK. (D) Current peak spacing $\Delta V_G$ in gate voltages. Here the results extracted from 14 measured consecutive current peaks are plotted for each case.

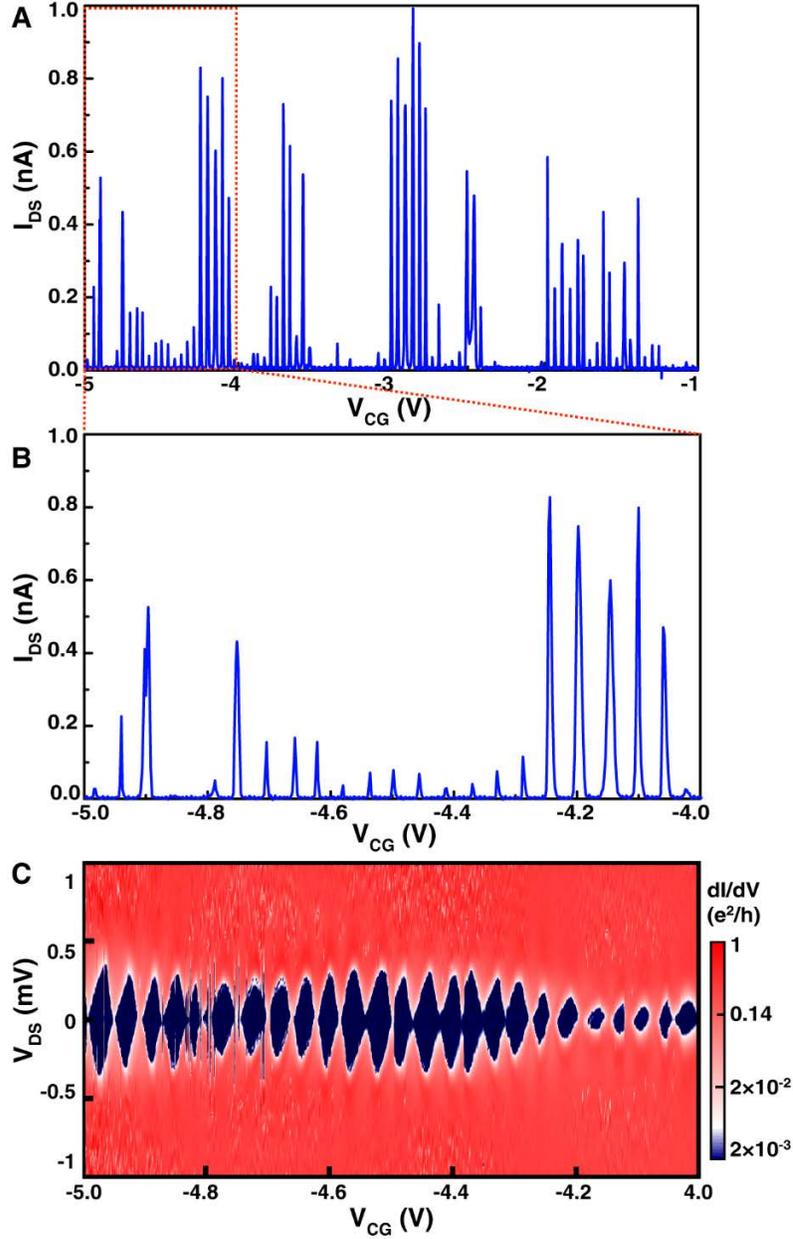

**Figure S7.** Tuning the transport properties of the $Bi_2Te_3$ single-electron transistor device (device #14) studied in the main article by the central gate voltage $V_{CG}$. (A) Source-drain current of the device measured as a function of $V_{CG}$ at $V_{DS}$ = 0.03 mV and $T$ = 40 mK. The upper, left constriction and right constriction gate voltages are set at $V_{UG}$ = -6.5 V, $V_{LG}$ = -3.6 V, and $V_{RG}$ = -3.3 V. (B) Zoom-in plot of the measurements in the region marked by the red-dashed rectangle in (A). (C) Differential conductance of the device measured as a function of $V_{DS}$ and $V_{CG}$ at $T$ = 40 mK in the same $V_{CG}$ range as in (B) and at $V_{UG}$ = -6.5 V, $V_{LG}$ = -3.6 V, and $V_{RG}$ = -3.3 V.

We have also measured the Coulomb current oscillations and the charge stability diagrams of the device (Device #14) by sweeping the central gate (CG) and the upper gate (UG) separately. Figure S7 shows the results of the measurements when the central gate (CG) is swept. Very similar to the measurements by sweeping the central and upper gates together (i.e., $V_{PG}$), sharp and roughly equally spaced Coulomb current oscillation peaks are observed in Figure. S7A,B. From the measurements, the peak spacing of $\triangle V_{CG} \sim 45$ mV can be extracted. The measurements by sweeping the upper gate show similar results. However, the peak spacing extracted from the measurements is smaller and is $\triangle V_{UG} \sim 15$ mV. This is consistent with the device structure in which the upper gate is coupled to the lithography defined $Bi_2Te_3$ Coulomb island much stronger than the central gate. Figure S7C shows the differential conductance as a function of source-drain bias voltage $V_{DS}$ and voltage $V_{CG}$ applied to the central gate only (charge stability diagram). Here, well defined Coulomb diamond regions of low differential conductance are again seen clearly.

Characteristics of the device measured after thermal cycling

To further verify the robustness of the Coulomb blockade effect seen in the device to further characterize the device, we have performed the measurements of the device after warming up the sample to room temperature, holding it for two days, and then cooling it down. Figure S8 shows the source-drain current of the device (Device #14) measured after the thermal cycling as a function of different gate voltages.

In Figure. S8A,B,D, it is seen that the Coulomb current peak structures are characteristically different from that in Figure S8C. In Figure. S8A,B,D, each broad current peak is superposed by small peaks (fine structures). We have estimated the small peak spacing in gate voltage and find $\triangle V_{CG} = 40$ mV, $\triangle V_{UG} = 15$ mV and $\triangle V_{RG} = 60$ mV [see the inset in Figure S8A]. The values of $\triangle V_{CG} = 40$ mV and $\triangle V_{UG} = 15$ mV are the same as that obtained before the thermal cycling. This is reasonable because these values reflect the capacitive coupling strengths of the central and upper gates to the Coulomb island defined by the FIB milling, and both the two gates and the Coulomb island should not be significantly changed in shape and size during the thermal cycling.

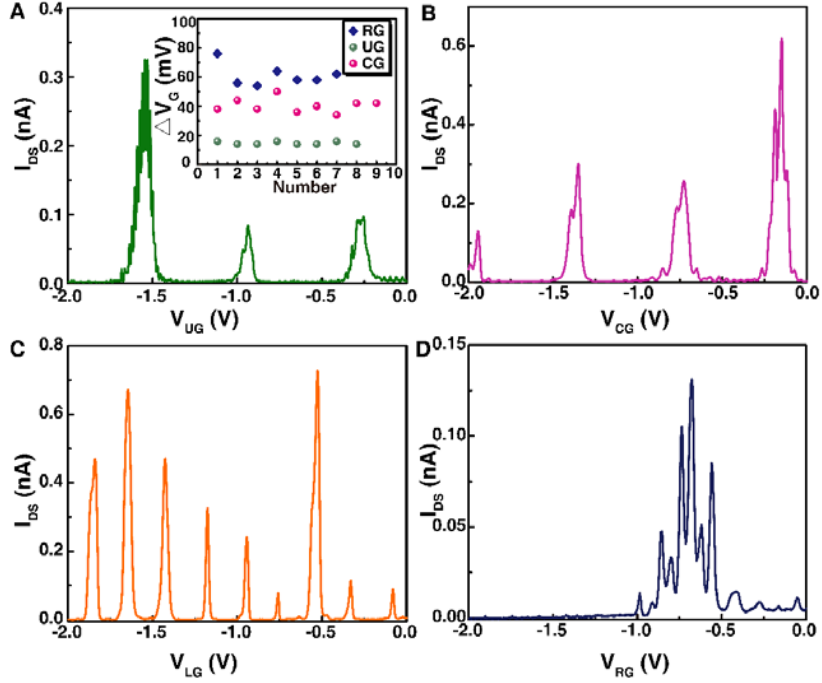

**Figure S8.** Characterization measurements of the $Bi_2Te_3$ single-electron transistor device (device #14) after a thermal cycling. (A) Source-drain current measured as a function of the upper gate voltage $V_{UG}$ at $V_{DS}$ = 0.1 mV and $T$ = 40 mK. The other gate voltages are set at $V_{LG}$ =$V_{RG}$ =$V_{UG}$ = 0 V in the measurements. (B) Source-drain current measured as a function of the central gate voltage $V_{CG}$ at $V_{DS}$ = 0.1 mV and $T$ = 40 mK. The other gate voltages are set at $V_{LG}$ =$V_{RG}$ = 0 V and $V_{UG}$ = -2 V. (C) Source-drain current as a function of voltage $V_{LG}$ applied to the left constriction gate at $V_{DS}$ = 0.1 mV and $T$ = 40 mK. The other gate voltages are set at $V_{UG}$ =$V_{RG}$ = $V_{CG}$ = 0 V. (D) Source-drain current of the device measured as a function of voltage $V_{RG}$ applied to the right constriction gate at $V_{DS}$ = 0.1 mV and $T$ = 40 mK. The other gate voltages are set at $V_{LG}$ =$V_{UG}$ =$V_{CG}$ = 0 V. The inset in (A) shows gate voltage spacings of fine Coulomb current oscillation peaks extracted from panels (A), (B) and (D).

In Figure S8C, Coulomb current peaks do not show superposed fine structures. These peaks are spaced by $\triangle V_{LG}$ = 220 mV. In comparison, we note that the broad current peaks seen in Figure. S8A,B are spaced by about 700 mV, which is much larger than the value of $\triangle V_{LG}$ = 220 mV seen in Figure S8C. Thus, the broad current peak structures can be traced to the presence of a small dot in the left constriction. Note that in Figure S8D, only one broad current peak is shown. This is because the gate voltage spacing of this peak to its neighboring peaks is large (larger than 1 V), in consistence with the fact that the broad peak originates from the same small dot located in the left constriction and the right constriction gate is located far from the dot.

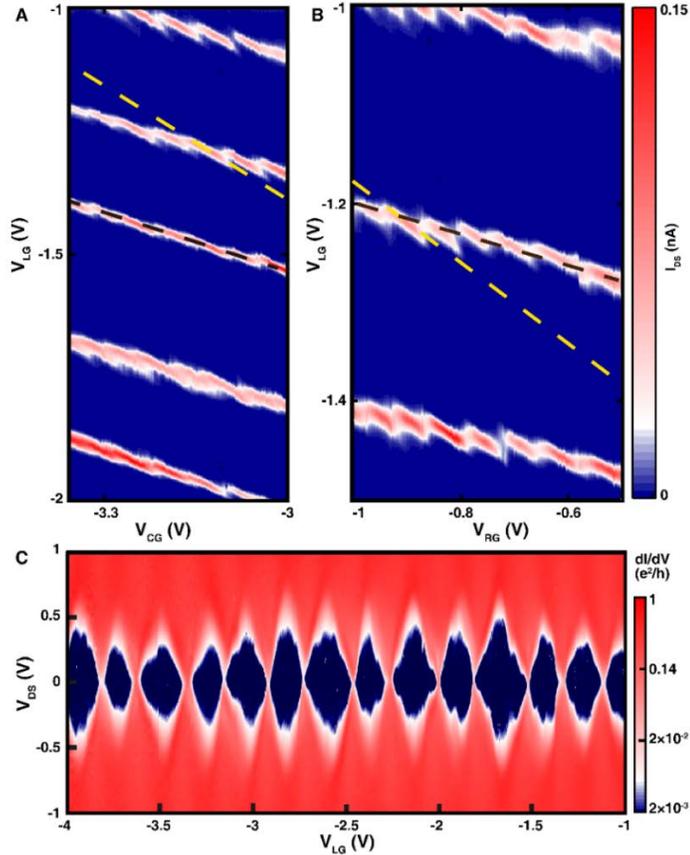

**Figure S9.** Additional characterization measurements of the $Bi_2Te_3$ single-electron transistor device (device #14) after the thermal cycling. (A) Source-drain current measured as a function of $V_{LG}$ and $V_{CG}$ at $V_{DS} = 0.035$ mV and $T = 40$ mK. The upper and right constriction gate voltages are set at $V_{UG} = -4.5$ V and $V_{RG} = 0$ V. (B) Source-drain current measured as a function of $V_{LG}$ and $V_{RG}$ at $V_{UG} = -4.5$ V and $V_{CG} = -3$ V. In (A) and (B), the current stripes with a slope marked by a yellow dashed line originate dominantly from single-electron tunneling through the central island, while the current stripes with a slope marked by a black dashed line originate dominantly from single electron tunneling through a dot in the left constriction area. (C) Differential conductance measured as a function of $V_{DS}$ and $V_{LG}$ at $T = 40$ mK. The other gate voltages are set at $V_{UG} = -4.5$ V, $V_{CG} = -3$ V and $V_{RG} = 0$ V in the measurements.

To further clarify the physical origins of the broad peaks and the fine structures, we show in Figure S9 the source-drain current of the device as a function of $V_{LG}$ and $V_{CG}$ [Figure S9A] and as a function of $V_{LG}$ and $V_{RG}$ [Figure S9B], and the differential conductance of the device as a function of source-drain bias voltage $V_{DS}$ and gate voltage $V_{LG}$ (charge stability diagrams) [Figure S9C] after the thermal cycling. In Figure S9A, two sets of current stripes are seen—a set of fine current stripes with the slope indicated by a yellow dashed line corresponding to the fine current peak structures seen in Figure S8 and a set of global current stripes with the slope indicated by a black dashed line corresponding to the broad current peak structures seen in Figure S8.

It can be identified from Figure. S9A,B that all the three gates—the left constriction gate, the right constriction gate and the central gate—can tune the fine current peak structures effectively, indicating that these fine current peaks originate from single-electron transport through a large dot located in between these gates, i.e., the central Coulomb island defined by the FIB milling. However, the global current peak structures can be tuned effectively only by the left constriction gate and thus these structures originate from a dot in the neighborhood of the gate—the left narrow constriction. Figure S9C shows the charge stability diagram of the dot. Here, clear and regular Coulomb diamonds are observed. From these Coulomb diamonds, the charging energy of the dot is extracted to be about 0.8 meV. This value is larger than that of the central Coulomb island, in consistence with the fact that the dot in the constriction is smaller than the central Coulomb island. Here we note that after the thermal cycling, although the right constriction still serves as a tunneling barrier for defining the central island, no signature of the formation of a dot in the right constriction is observed.

Overall, the fine current peak structures seen in Figures. S8 and S9 originate from the single electron tunneling through the central island defined by the FIB milling and the broad current peak structures shown in Figures. S8 and S9 originate from the single electron transport through a small dot located in the left constriction. It is also shown that the fine Coulomb current oscillation peaks are suppressed between slowly oscillated, broad Coulomb current peaks, indicating that the constriction is less transparent in the Coulomb blockade regions of the small dot in the constriction. Clearly, the potential fluctuation generated localized dot may change in size, shape, and position during the thermal cycling process. The central Coulomb island defined by the FIB milling and its single-electron transport properties should survive and remain unchanged after the thermal cycling.

## Measurements of Additional Bi$_2$Te$_3$ Single-electron Transistor Devices

Two other Bi$_2$Te$_3$ single-electron transistor devices made by the same FIB milling process are measured at low temperatures. Figure S10 shows the measurements of a device made on the same chip as device #14 but from a different nanoplate. As shown in Figure S10A, the device possesses a circular dot and two side gates, labelled by PG, placed symmetrically around the dot. The thickness of the nanoplate is 8 nm, the dot radius is 80 nm, and the two narrow constrictions have a length of 20 nm and a width of 50 nm. Figure S10B shows clear Coulomb current oscillations and Figure S10C shows the charge stability diagram of the device. The single-electron charging energy of the dot extracted from the charge stability diagram is about 5 meV or larger. This is consistent with the fact the dot is much smaller in both lateral size and thickness.

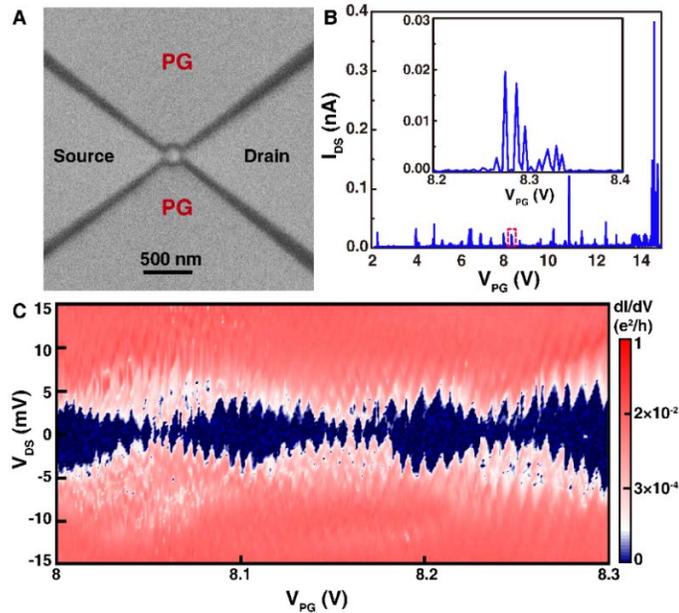

**Figure S10.** Measurements of a different Bi$_2$Te$_3$ single-electron transistor device (device #33) made of a Bi$_2$Te$_3$ nanoplate grown on the same mica substrate as the nanoplate used for device #14. (A) SEM image of the device measured. The device is made of a Bi$_2$Te$_3$ nanoplate of 8 nm in thickness and consists of a circular dot with a radius of 80 nm, two narrow constrictions with a width of 50 nm and a length of 20 nm, and two side gates. (B) Source-drain current measured as a function of voltage $V_{PG}$ applied to the two side gates at source-drain voltage $V_{DS}$ = 2 mV and temperature $T$ = 40 mK. The inset shows zoom-in plot of the measurements in the region marked by a red rectangle in the figure. (C) Differential conductance of the device as a function of $V_{PG}$ and $V_{DS}$ (charge stability diagram) at $T$ = 40 mK.

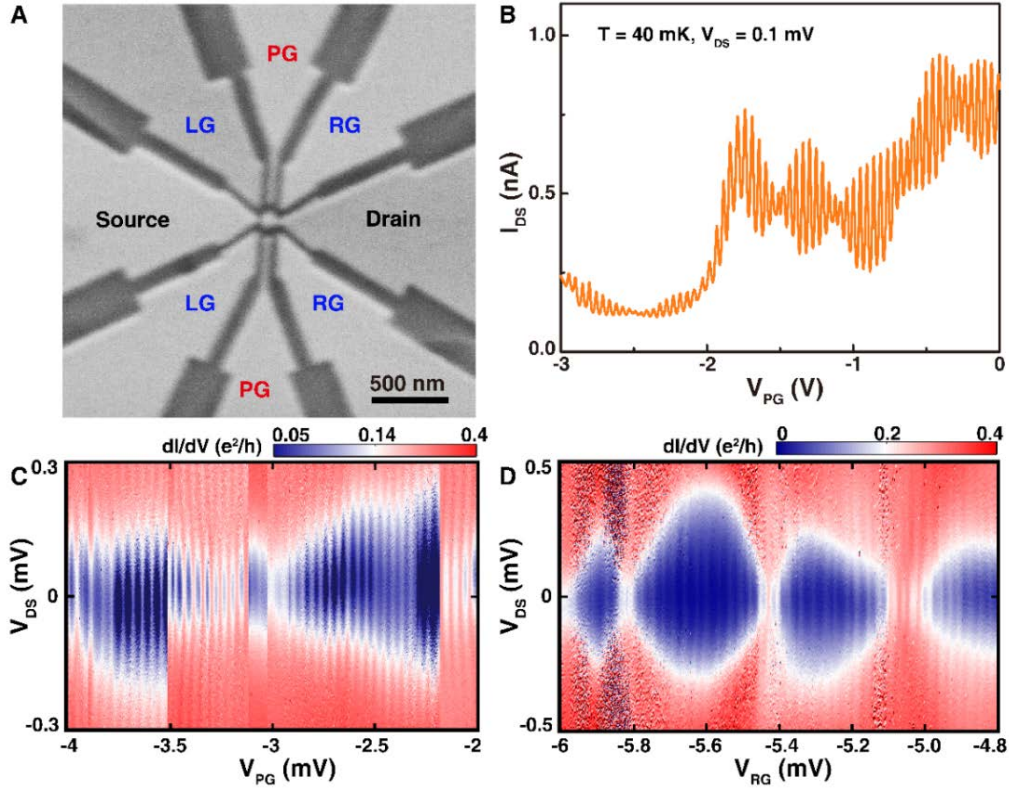

**Figure S11.** Measurements of a single-electron transistor device (device #5) made from a $Bi_2Te_3$ nanoplate grown on another mica substrate. (A) SEM image of the device measured. The device is made of a $Bi_2Te_3$ nanoplate of 14 nm in thickness and consists of a circular dot with a radius of 110 nm, two narrow constrictions with a width of 40 nm and a length of 25 nm, and six side gates. (B) Source-drain current of the device measured as a function of voltage $V_{PG}$ applied to the two central plunger gates at source-drain voltage $V_{DS}$ = 0.1 mV and temperature $T$ = 40 mK. (C) Differential conductance measured as a function of $V_{PG}$ and $V_{DS}$ at $T$ = 40 mK. The two left constriction and the two right constriction gate voltages are set at $V_{LG}$ = -1.11 V and $V_{RG}$ = -1.11 V in the measurements. (D) Differential conductance measured as a function of $V_{RG}$ and $V_{DS}$ at $T$ = 40 mK. The two left constriction and the two central plunge gate voltages are set at $V_{LG}$ = -8.2 V and $V_{PG}$ = -8 V in the measurements.

Figure S11 shows the measurements of a device made by the FIB milling process on a different chip. As shown in Figure S11A, the device consists of a circular dot with a radius of 110 nm, two narrow constrictions with a width of 40 nm and a length of 25 nm, and six side gates with two (labelled by LG) located near the left narrow constriction, two (labelled by PG) in the middle, and two (labelled by RG) near the right constriction. The nanoplate employed for making this device has a thickness of 14 nm. Figure S11B shows the measured source-drain current as a function of voltage $V_{PG}$ applied to the two PG gates. Here, fine fast Coulomb oscillations are seen to superimpose on the background oscillated current. The finite background current seen in the measurements indicates that the central island is in a relatively open regime where the two confinement barriers are rather transparent. Figure S11C shows the differential conductance of the device as a function of source-drain voltage $V_{DS}$ and voltage $V_{PG}$ applied to the two PG gates (charge stability diagram). In this figure, the Coulomb diamonds corresponding to the fine fast current oscillations seen in Figure S11B are clearly observable. It is also seen that these Coulomb diamonds are modulated by the transparency of the constrictions. The single-electron charging energy extracted from the charge stability diagram is about 0.3 meV. Figure S11D shows the differential conductance of the device as a function of source-drain voltage $V_{DS}$ and voltage $V_{RG}$ applied to the two RG gates. Here, small Coulomb diamond structures are also clearly visible, implying that the electron number in the central island can be sensitively tuned by the two right gates.

## Measurements of single constrictions

To demonstrate the width-dependent transport properties of narrow constrictions, we have measured transport characteristics of single constrictions with different widths. The inset of Figure S12B shows the structure of these single constriction devices, where the constriction length is fixed at 25 nm and the constriction width varies from 50 nm to 400 nm. The nanoplates used for making these constriction devices are especially selected and have a thickness of about 10 nm, as indicated in Figure S12B. Figure S12A shows the resistance of each nanoplate measured at room temperature before and after a constriction is fabricated by FIB milling. It is seen that the resistance of a nanoplate increases by one order of magnitude after FIB milling out the constriction from it, as a result of a substantial reduction in conduction channel width. It is also seen that the resistance increases with decreasing constriction width. Figure S12B shows the conductance of five devices with different constriction widths measured at $T = 2$ K. For the device with a constriction width of 50 nm, the conductance at $T = 2$ K is about one quantum conductance. Figure S12C shows temperature dependent measurements of the resistance of four of these five devices. The temperature dependences of the resistance of the devices with a constriction width of 150, 250, or 400 nm decrease with decreasing temperature, which are similar to that of a $Bi_2Te_3$ nanoplate, showing a metallic behavior. In contrast, the device with a constriction width of 50 nm behaves very differently; its resistance increases as temperature decreases, showing a semiconductor behavior. Figure S12D shows the Arrhenius plot of $\ln(R)$ versus $1/T$ for the device with 50 nm in constriction width.

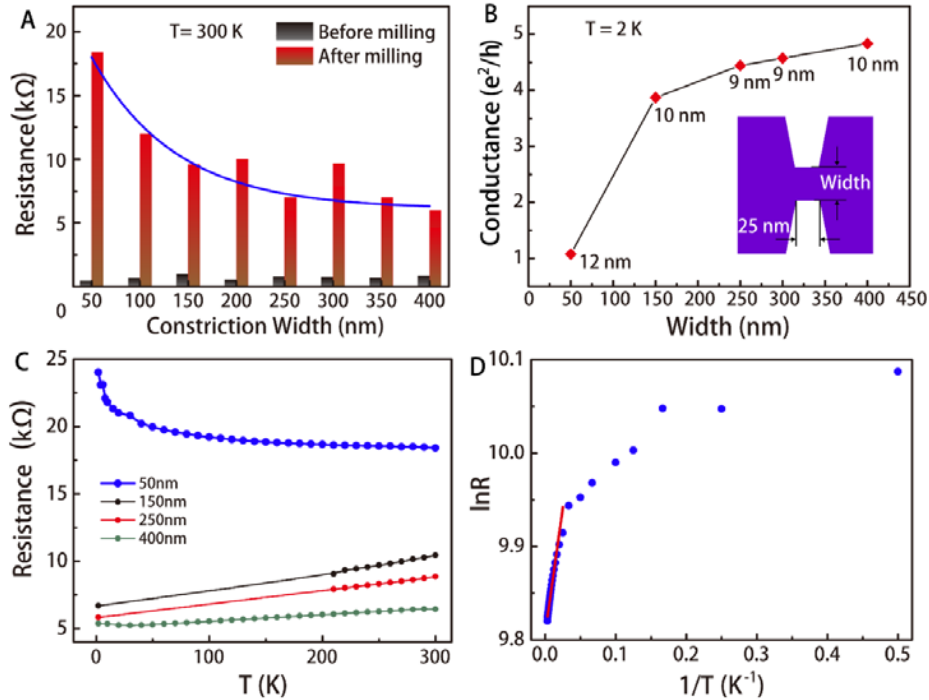

**Figure S12.** Transport characteristics of single-constriction devices, as schematically shown in the inset of (B), with a 25-nm constriction length but different constriction widths made from $Bi_2Te_3$ nanoplates with ~10 nm in thickness. (A) Resistance of a few nanoplates measured at room temperature before and after single constrictions of different widths are made by FIB milling. The blue solid line is guide to the eyes. (B) Conductance of five single-constriction devices with different constriction widths at $T = 2$ K. The thickness of each nanoplate from which the device is made is marked beside the corresponding data point. (C) Temperature dependences of the resistance of the devices with constriction widths of 50 nm, 150 nm, 250 nm, and 400nm. (D) Arrhenius plot of the resistance of the device with 50 nm in constriction width versus the inverse of temperature. The red solid line is a line fit to the high temperature data.